\documentclass[galaxies,article,accept,moreauthors,pdftex]{Definitions/mdpi}
\usepackage{natbib}

\newcommand{\nat}{Nature}      
\firstpage{1}
\pubvolume{9}
\issuenum{4}
\articlenumber{97}
\pubyear{2021}
\copyrightyear{2021}
\externaleditor{{Academic Editors: Francesca Loi and Tiziana Venturi}} 
\datereceived{6 October 2021}
\dateaccepted{2 November 2021}
\datepublished{8 November 2021}
\hreflink{https://doi.org/10.3390/~\linebreak~galaxies9040097} 
\pdfoutput=1

\Title{A Multiwavelength Dynamical State Analysis of ACT-CL J0019.6+0336}

\TitleCitation{A Multiwavelength Dynamical State Analysis of ACT-CL J0019.6+0336}

\Author{Denisha S. Pillay 
$^{1,2,}$*\orcidA{}, {David J. Turner} 
$^{3}$\orcidB{}, Matt Hilton $^{1,2}$\orcidC{}, Kenda Knowles $^{4,5}$\orcidD{}, Kabelo C. Kesebonye $^{1,2}$, Kavilan Moodley $^{1,2}$\orcidE{}, Tony Mroczkowski $^{6}$\orcidF{}, Nadeem Oozeer $^{7,8}$\orcidG{}, Christoph Pfrommer $^{9}$\orcidJ{}, \mbox{Sinenhlanhla P. Sikhosana} $^{1,2}$\orcidK{} and Edward J. Wollack $^{10}$\orcidL{}}

\AuthorNames{Firstname Lastname, Firstname Lastname and Firstname Lastname}

\AuthorCitation{Pillay, D.S.; Turner, D.J.; Hilton, M.; Knowles, K.; Kesebonye, K.C.; Moodley, K.;  Mroczkowski, T.; Oozeer, N.; Pfrommer, C.; Sikhosana, S.P.; et al.
}

\address{%
$^{1}$ \quad Astrophysics Research Centre, University of KwaZulu-Natal, Durban 4041, South Africa; hiltonm@ukzn.ac.za~(M.H.); kablokes@gmail.com (K.C.K.); kavilan.moodley@gmail.com (K.M.); sikhosanas@ukzn.ac.za (S.P.S.)
\\
$^{2}$ \quad School of Mathematics, Statistics \& Computer Science, University of KwaZulu-Natal, Westville Campus, 
Durban 4041, South Africa\\
$^{3}$ \quad Department of Physics and Astronomy, Pevensey Building, University of Sussex, Brighton BN1 9QH, UK; david.turner@sussex.ac.uk\\
$^{4}$ \quad Department of Physics and Electronics, Rhodes University, P.O. Box 94, Makhanda 6140, South Africa; kendaknowles.astro@gmail.com\\
$^{5}$ \quad South African Radio Astronomy Observatory, 2 Fir Street, Observatory, Cape Town 7405, South Africa\\
$^{6}$ \quad European Southern Observatory, Karl-Schwarzshild-Str. 2, D-85748 Garching b. München, Germany; tony.mroczkowski@eso.org\\
$^{7}$ \quad South African Radio Astronomy Observatory, 2 Fir Street, Black River Park, Observatory, \mbox{Cape Town 7925, South Africa}; oozeern@gmail.com\\
$^{8}$ \quad African Institute for Mathematical Sciences, 6 Melrose Road, Muizenberg 7945, South Africa\\
$^{9}$ \quad Leibniz-Institut für Astrophysik Potsdam (AIP), An der Sternwarte 16, 14482 Potsdam, Germany; cpfrommer@aip.de\\
$^{10}$\quad NASA/Goddard Space Flight Centre, Greenbelt, MD 20771, USA; edward.j.wollack@nasa.gov
}

\corres{Correspondence: denishapillay5@gmail.com}

\abstract{
In our study, we show a multiwavelength view of ACT-CL J0019.6+0336 (which hosts a radio halo), to investigate the cluster dynamics, morphology, and ICM. We use a combination of XMM-Newton images, Dark Energy Survey (DES) imaging and photometry, SDSS spectroscopic information, and 1.16 GHz MeerKAT data to study the cluster properties. Various X-ray and optical morphology parameters are calculated to investigate the level of disturbance. We find disturbances in two X-ray parameters and the optical density map shows elongated and axisymmetric structures with the main cluster component southeast of the cluster centre and another component northwest of the cluster centre. We also find a BCG offset of $\sim$950 km/s from the mean velocity of the cluster, and a discrepancy between the SZ mass, X-ray mass, and dynamical mass ($M_{X,500}$ and $M_{SZ,500}$ lies $>3\sigma$ away from $M_{\rm{dyn},500}$), showing that J0019 is a merging cluster and probably in a post-merging phase.
}

\begin{document}

\section{Introduction}\label{sec1}
Multiwavelength observations of galaxy clusters carry an abundance of information, giving insight into the intracluster medium (ICM), its thermal and non-thermal components, and the cluster dynamics. Galaxy clusters merge with neighbouring clusters through the most energetic events in the Universe since the big bang~\citep{2007A&A...474..745P}. The merger disturbs the cluster's natural state through merger-driven turbulence and often leave imprints on the~ICM.

Turbulence in the ICM also affects individual sources in the cluster, particularly radio galaxies~\citep{2015IAUS..313..315B}. There are differences in radio emission from cluster galaxies and isolated galaxies that are generally attributed to the interaction of the emission with the ICM. Most radio emissions from galaxies have a compact radio source associated with the active galactic nuclei (AGN) and extended regions of radio emission (radio lobes) quite distant from the compact radio source. The morphology of extended radio emission associated with individual galaxies (such as bent tail galaxies) is strongly influenced by the environment in which the galaxy exists~\citep{10.1046/j.1365-8711.2000.03079.x}. The motion of galaxy clusters through the dense cluster gas is widely recognized as the mechanism which produces bent radio source tails and the dynamic pressure responsible for a variety of other observed source shapes~\citep{10.1046/j.1365-8711.2000.03079.x}.

A fraction of the gravitational energy released during violent merger events is converted into magnetic-field amplification and results in the acceleration of high energy particles in the ICM, giving rise to diffuse radio sources~\citep{2012A&ARv..20...54F}. Cluster-scale diffuse radio sources have steep spectra, exhibit low surface brightness, and are broadly categorized into three groups: radio halos, radio mini halos, and radio relics. Radio halos and mini halos are centrally located sources, whereas relics are found as filamentary, elongated shapes located at the cluster periphery (see~\citep{2019SSRv..215...16V} for a recent review).

Diffuse radio emission and multiwavelength studies provide more knowledge about the link between the underlying thermal and non-thermal processes present in the ICM~\citep{2014IJMPD..2330007B}. Several open questions regarding cluster diffuse emission exist. It is not understood why some merging clusters host radio halos and relics while others do not show any evidence of extended radio emission ~\citep{2010ApJ...721L..82C}. The correlation between non-thermal diffuse emission and the dynamical properties of the host clusters is unclear and the underlying formation mechanisms are not well understood~\citep{2010ApJ...721L..82C}. These open questions have larger implications on topics and processes which include dark matter, AGN feedback mechanisms, and the large-scale structure of the universe. The sensitivity of new generation radio telescopes will enable studies of large and diverse samples of clusters at higher redshift and lower mass.

Previous studies of radio diffuse emission were limited to high mass and low redshift samples, and clusters with high redshifts were generally solitary detections~\citep{2014ApJ...786...49L,2016MNRAS.459.4240K}.
Large sample studies, such as~\citet{Cuciti_2021}, are crucial to a complete understanding of the evolutionary life of radio diffuse emission, formation theories of radio sources, and analysis of radio source statistics. The MeerKAT Exploration of Relics, Giant Halos and Extragalactic Radio Sources (MERGHERS;~\citep{2016mks..confE..30K}) 
project aims to be a mass-selected sample of $\sim$200 Sunyaev--Zel'dovich (SZ) selected clusters. These clusters will cover a suitable range of criteria to ensure an unbiased study. In preparation for such a large project, a pilot sample of 13 clusters has been obtained using the MeerKAT telescope~\citep{2021MNRAS.504.1749K}.

In this paper, we perform a multiwavelength follow-up of ACT-CL J0019.6+0336 (hereafter J0019), presented in~\citet{2021MNRAS.504.1749K}, to further understand the dynamics of the cluster.
The paper is structured as follows. In Section \ref{sec2}, we show existing multiwavelength data on J0019. Section \ref{sec3} shows the morphology analysis methods and results, followed by the discussion and conclusion in Sections \ref{sec4} and \ref{sec5} respectively.
In this paper, we adopt a $\Lambda$CDM flat cosmology with $H_0$ = 70 kms$^{-1}$ Mpc$^{-1}$ , $\Omega_\Lambda$ = 0.7,  $\Omega_m$ = 0.3.



\section{Observations and Data}\label{sec2}
J0019 lies at $z=0.266$ and was first detected in a catalogue of galaxy clusters derived by~\citet{1965cgcg.book.....Z} (ZwCl 0017.0+0320) 
and was later observed by the Atacama Cosmology Telescope (ACT;~\citep{2018ApJS..235...20H}). 
In this section, we describe the multi-wavelength data and MeerKAT observations used for this study. The relevant cluster properties are given in Table \ref{tab11}.

\begin{specialtable}[H]
\caption{Published properties of J0019. The position, redshift, SZ mass, and radio properties are from 
~\citet{2021MNRAS.504.1749K}, with the mass being the weak lensing-calibrated value. Integrated 0.5--2.0~keV X-ray luminosity is calculated within $R_{500}$ = 1389.5 kpc (spherical radius that encloses an average density equal to 500 times the critical density at the cluster redshift, z = 0.266).\label{tab11}}
\setlength{\tabcolsep}{22mm}\begin{tabular}{cc}
\toprule

R.A.$_{J2000}$ (deg) & 4.91085\\
Dec.$_{J2000}$ (deg) & 3.60879\\
redshift & 0.266\\
$M_{\rm 500c, SZ}$ ($10^{14} M_\odot $ )&10.2 $\pm$ 2.28\\
$L_{500,X} (10^{44}$ ergs/s) & 4.94 $\pm$ 0.032\\
$S_{\rm 1.16 GHz}$ (mJy)& 9.16 $\pm$ 0.57\\
log$P_{1.4GHz}$ (W/Hz)& 24.13 $\pm$ 0.06\\
\bottomrule
\end{tabular}
\end{specialtable}

%

\subsection{Millimetre}\label{sec2.1}
Studies such as~\citet{2012MNRAS.421L.112B} and~\citet{2015A&A...580A..97C}, have shown that samples selected via their SZ signal show a higher detection rate than X-ray selected samples and SZ selected samples have the benefit of the flux limit translating directly into a mass limit~\citep{2014MNRAS.437.2163S}. Various SZ telescopes now provide large SZ-detected cluster samples in which galaxy clusters are detected via the distortion of the cosmic microwave background (through inverse Compton scattering of high-energy electrons).
{}
J0019 was observed in both Planck and ACT surveys~\citep{2021ApJS..253....3H,2015A&A...581A..14P}. We make use of the ACT observations because they have higher sensitivity and resolution than the Planck observations.

J0019 is part of the fifth release of the ACT cluster catalogue (\citep{2018ApJS..239...18A}, ACT DR5) 
which contains more than 4000 SZ selected optically confirmed clusters~\citep{2021ApJS..253....3H}. J0019 was observed as part of the larger SZ-selected sample from ACT that forms the targets for MERGHERS with a signal-to-noise ratio, S/N = 25.23 (at the reference 2.4' filter scale)~\citep{2021ApJS..253....3H}. This shows that J0019 is a very massive and energetic cluster.

\subsection{Optical}\label{sec2.2}
We make use of available Dark Energy Survey (DES) DR1 photometry for J0019. \textsc{zCluster}\endnote{\url{https://github.com/ACTCollaboration/zCluster} (accessed on 1 June 2021)
.} estimates the photometric redshift of a particular galaxy cluster using multi-band optical and infrared photometry. The optical photometry is obtained using the \textsc{zCluster} code and
the photometric redshift of the cluster is estimated by the weighted sum of individual galaxy probability distributions in the direction of each galaxy~\citep{2018ApJS..235...20H}. These probability distributions are found using a template fitting method as seen in~\mbox{\citet{2000ApJ...536..571B}}, where a set of default galaxy spectral energy distribution (SED) templates~\citep{2021ApJS..253....3H} are fitted to the observed broadband SED of each galaxy.
Moreover, spectroscopic observations are available for 17 possible cluster members (within $R_{500}$) in the Sloan Digital Sky Survey (SDSS) DR16 database~\citep{2020ApJS..249....3A}.

\subsection{Radio}\label{sec2.3}
\begin{figure}[H]
\widefigure
\centering
\includegraphics[height=6.5cm]{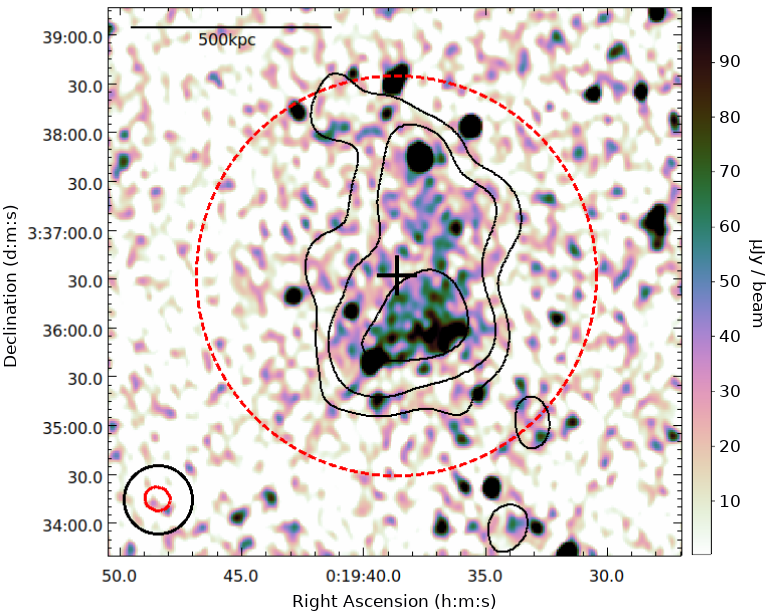}\hspace{0.5cm}
\includegraphics[height=6.5cm]{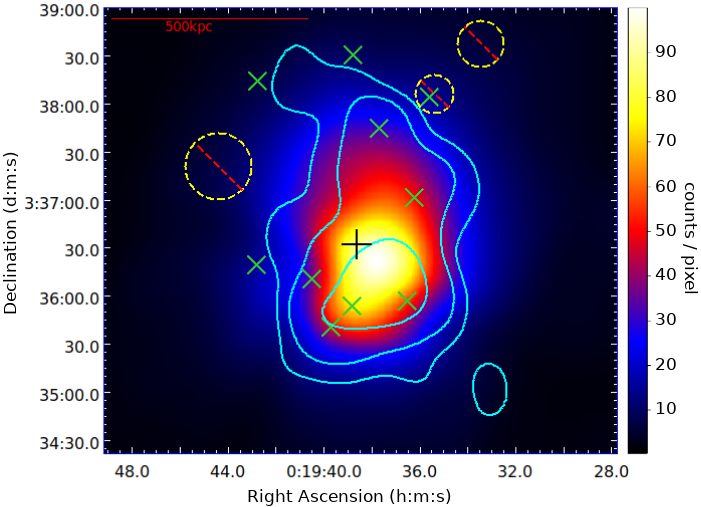}
\caption{\textbf{Left:} Full-resolution MeerKAT L-band radio image for J0019 with large scale filtered image contours overlaid. Contours are at [$-$3, 3, 5, 10] × $\sigma$ (52 $\upmu$Jy/beam) of the filtered image. The synthesized beam for both the full-resolution (red ellipse, 7.9$^{\prime\prime}$ $\times$ 7.2$^{\prime\prime}$, p.a. 162$^{\circ}$) and filtered map (black circle, 21$^{\prime\prime}$) are indicated in the lower left of the panel. The dashed red circle denotes $R_{500 kpc}$ centred on the ACT SZ peak, which is marked by a black cross. Full resolution central rms noise is 18.1 $\upmu$Jy/beam. \textbf{Right:} XMM-Newton combined PN, MOS1, and MOS2 images of J0019 with the large-scale radio contours overlaid. The black cross represents the ACT SZ peak and the green Xs are centred on the peak radio emission of sources found in the MeerKAT radio image. Regions of the image masked after removing X-ray point sources are indicated by yellow, dashed, excluded circles. The image is unbinned and has been smoothed by a Gaussian with a kernel radius of 6~pixels (1 px = 4.35$^{\prime\prime}$).  \label{sfig1}}
\end{figure}

J0019 has been observed by the MeerKAT telescope for $\sim$24 min at L-band~\linebreak~(\mbox{$\sim$0.9--1.7~GHz}) (\citep{2021MNRAS.504.1749K}, PI:Knowles).
First, second, and third-generation MeerKAT calibration algorithms have been applied to the radio data to correct for direction independent and dependent effects. First and second-generation calibration was applied to the field using the \texttt{oxkat} v1.0\endnote{\url{https://github.com/IanHeywood/oxkat} (accessed on 1 June 2020.)
} pipeline~\citep{2020ascl.soft09003H}, a Python-based reduction pipeline for MeerKAT. Third-generation calibration has been applied to the field using
\texttt{DDFACET}\endnote{\url{https://github.com/saopicc/DDFacet} (accessed on  1 July 2020).
} and
\texttt{Killms}\endnote{\url{https://github.com/saopicc/killMS} (accessed on 1 July 2020).
} to correct for corrupting artifacts in the radio image. The
\texttt{KATBEAM}\endnote{\url{https://github.com/ska-sa/katbeam} (accessed on 1 August 2020).
} package is used to create a primary beam corrected final image for
analysis. A large scale filtered image (contours overlaid in Figure \ref{sfig1}) is also created through an image-plane filtering technique introduced by~\mbox{\citet{2004JKAS...37..329R}}, in which extended
emission without compact sources is highlighted by filtering out emission on a scale 1--3 times the synthesised beam. The left panel of \mbox{Figure~\ref{sfig1}} shows the final full-resolution radio map of the J0019 cluster region with large-scale filtered image contours overlaid. A radio halo ($\sim$810 kpc in size) has been detected with a 1.16\,GHz flux density of $9.16 \pm 0.57$ mJy. The power value has been estimated as \mbox{$P_{\rm 1.4GHz}$ = $24.13 \pm 0.06$~W/Hz} by using a fiducial spectral index, $\alpha = -1.3 \pm 0.4$ and extrapolating the MeerKAT flux densities to 1.4 GHz (adopting the spectral power law convention of S$_\nu$ $\propto$ $\nu^{\alpha}$).

\begin{figure}[H]
\widefigure
\includegraphics[width=0.9\textwidth]{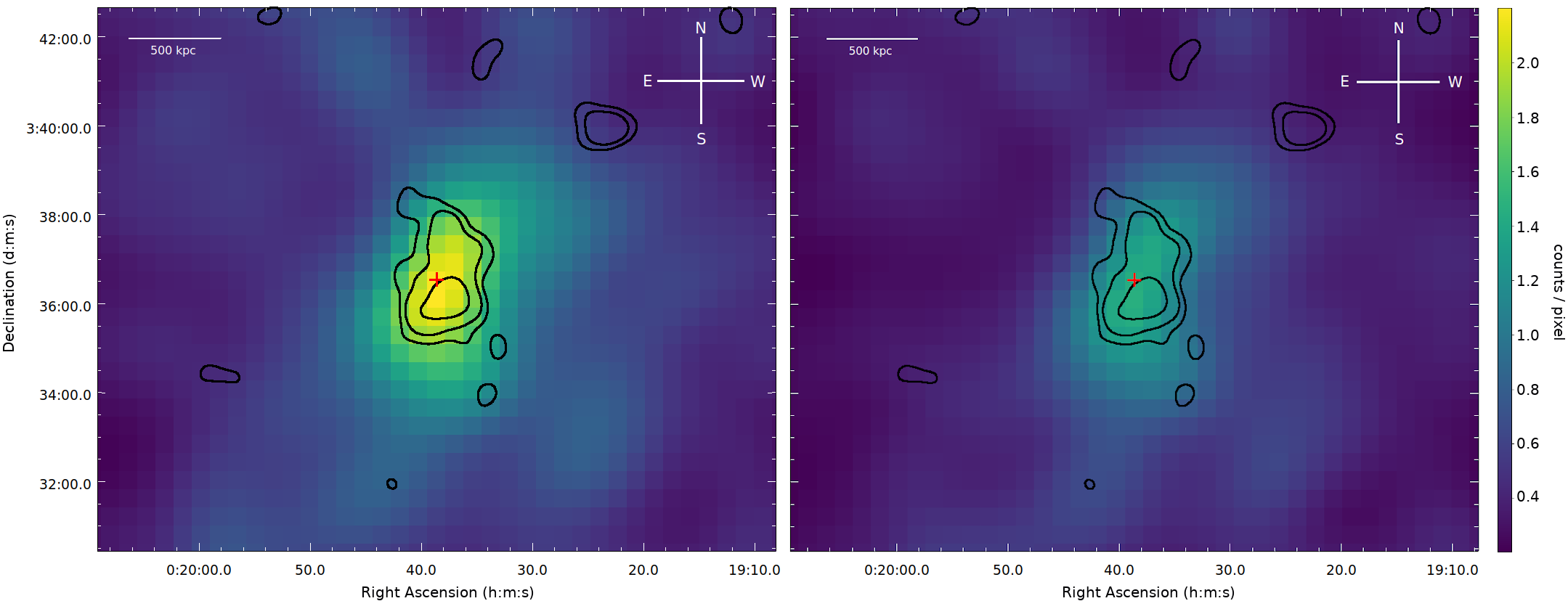}
\caption{\textbf{Left:} Density map of J0019 obtained from \textsc{zField} using photometry from the DES DR1 database and z = 0.266. The projected density map is 4 Mpc $\times$ 4 Mpc with each pixel 0.2 Mpc $\times$ 0.2 Mpc in size. Black large-scale radio contours are overlaid and the red cross indicates the SZ peak of the cluster. The units of the colour scale are counts per pixel. \textbf{Right:} Corresponding error map for the J0019 galaxy density map.\label{fig3}}
\end{figure}

\end{paracol}
\nointerlineskip
\begin{paracol}{2}
\switchcolumn

\vspace{-10pt}
\subsection{X-ray}\label{sec2.4}
We use archival data from the XMM-Newton space telescope to provide an X-ray view of J0019 and probe its morphology. The only XMM data available for J0019 was obs ID: 0693010301, a pointed, 47.9\,ks observation of the cluster taken in 2012. Events from all three EPIC (PN, MOS1, and MOS2)\endnote{
The XMM-Newton observations include European Photon Imaging Camera (EPIC) data from the two MOS (Metal Oxide Semi-conductor) CCD arrays and the pn CCD array.} cameras have been used.
The XMM Cluster Survey~(\citep{Romer_2001}, XCS) 
cleaned the data and performed source detection with a custom version of WAVDETECT~\citep{wavdetect} (called the XCS Automated Pipeline Algorithm (XAPA)). XAPA is run on merged EPIC (PN+MOS1+MOS2) images, with a pixel size of 4.35$^{\prime\prime}$, that have been generated using events within an energy range of 0.5--2.0 keV. Once sources in an image have been located, XAPA classifies them as either point or extended.

We use {\em XMM}: Generate and Analyse (\texttt{XGA}\endnote{\href{https://github.com/DavidT3/XGA}{{\em XMM}: Generate and Analyse GitHub} (accessed on 1 May 2021).
}; \citep{xgapaper}), 
a new open-source Python module created for XCS, to generate soft-band (0.5--2.0 keV) count-rate maps. \texttt{XGA} then uses the XCS source catalogue to produce a mask to remove irrelevant sources. It identifies the XCS source that matches J0019, then masks all other sources. The final processed X-ray image is shown in the right panel of Figure~\ref{sfig1}.

The unabsorbed X-ray luminosity in the 0.5--2.0 keV energy band was measured by performing simultaneous fits to spectra generated within $R_{500}$\endnote{Spherical radius that encloses an average density equal to 500 times the critical density at the cluster redshift, z = 0.266.} of J0019, centred on the~\citet{2021MNRAS.504.1749K} coordinates. We fit absorbed (with \texttt{tbabs}, \cite{tbabs}) 
plasma emission models (APEC, \citep{apec})
to the three spectra, with temperatures linked, metallicity fixed at 0.3~$Z_{\odot}$, redshift fixed at 0.266, and nH fixed at $0.029\times10^{22}$ cm$^{-2}$ (taken from the full-sky HI survey by the \cite{nh}). The fit is performed using \texttt{XGA}'s XSPEC \citep[][]{xspec} interface, SAS v17.0.0, and XSPEC v12.10.1. We measure a temperature of $T = 6.7 \pm 0.1$ keV within $R_{500}$ and a bolometric X-ray luminosity of $L_X = (1.86 \pm 0.02) \times 10^{45}$ ergs/s. Comparing to the M$_{x}$-T$_{x}$ relation in \cite{2006ApJ...641..752K}, we estimate a mass within R$_{500}$, M$_X\sim$~(6.5$~ \pm ~1.3)\times10^{14}M_\odot$.

\section{Cluster Morphology}\label{sec3}
Galaxy clusters form via a hierarchical sequence of accretion and mergers of smaller interacting substructures~\citep{2014IJMPD..2330007B}. Current observations of radio diffuse emission favour theories of merger-driven radio halo and radio relic formations~\citep{2019SSRv..215...16V}. Therefore, it is important to understand the dynamical state of J0019. Numerous studies, e.g.,~\citet{2010A&A...514A..32B,2013MNRAS.436..275W} have analyzed the substructure of clusters to determine their dynamical states using multiwavelength data. By studying the form or shape of the cluster we can infer the dynamics. We use available multiwavelength information to estimate a variety of X-ray and optical-derived parameters.
\subsection{Optical Morphology}\label{sec3.1}
\textsc{zCluster} retrieves and uses redshift probability distributions to measure a galaxy cluster's photometric redshift. Thereafter, we create a projected 2-D density map of the galaxy cluster using the right ascension (in degrees), declination (in degrees), photometric redshift of the cluster, and a catalogue of redshift probability distributions of individual galaxies. We use the right ascension and declination of each galaxy to position them onto a 2-D map as done in~\citet{2013MNRAS.436..275W}. The galaxy coordinates are converted into Cartesian coordinates and centred at the input R.A. and Dec. (SZ peak), on a projected \mbox{4 Mpc $\times$  4~Mpc} map. The map is made up of 20 $\times$ 20 bins, each with 0.2 Mpc length. The positions of the galaxies in Cartesian coordinates are obtained by finding the angular distance between the coordinates of individual galaxies and the centre of the cluster.

The peak of the probability distribution gives the maximum likelihood of a galaxy having a specific redshift~\citep{2018ApJS..235...20H}. We obtain a weighting for our plot by integrating the probability distribution of each galaxy around the photometric redshift (bounds of integration from z\_phot - delta\_z to z\_phot + delta\_z).
The \textsc{zCluster} code takes the above-calculated quantities and produces a 2-D projected density map of the galaxy cluster. Thereafter, the map is convolved with a Gaussian kernel (with smoothing length = 2 pixels) to smooth the~map.

A density uncertainty map (right panel of Figure \ref{fig3}) is created using Monte Carlo simulations. We add Gaussian-distributed noise to the galaxy photometry at each Monte Carlo step and recompute the photometric redshifts, probability distributions, and density maps. After 1000 iterations, the 68th percentile of the maps is taken as an estimate of the error map.

\end{paracol}
\nointerlineskip
\begin{paracol}{2}
\switchcolumn

The density map for J0019 (left panel of Figure~\ref{fig3}) is made using zField (a subclass of zCluster) in which we have used photometric redshift, z = 0.266, and a redshift range, delta\_z = 0.2. The weighting for this density map is obtained by integrating the probability distribution for each galaxy around z = 0.266 (bounds of integration from 0.266 $-$ 0.2 to 0.266 + 0.2).

\subsubsection{Centre Shift}\label{sec3.1.1}
The centre shift method used in this work is a simple method to determine the dynamical state of a cluster by calculating the offset between the SZ peak and the peak in the optical density map. The centre of the cluster generally hosts the brightest cluster galaxy (BCG). In relaxed clusters, the centre of galaxy clusters traced by the gas using X-ray, or the SZ effect, coincides (within a few kpc) with the centre of the galaxy cluster traced optically. This can be contrasted with the centres of mergers (determined using X-ray and SZ data), which can be hundreds of kpc away from the centroid determined optically as seen in~\citet{2019MNRAS.482.5093P}.

We measure the distance between the coordinates of the maximum of the density plot and the coordinates of the cluster SZ peak~\citep{2001MNRAS.320...49K}.
After finding the coordinates of the peak value of the density plot (left panel of Figure~\ref{fig3}), we use the distance Formula (\ref{equ}) to find the centre shift.
\begin{equation}
\label{equ}
CS=\sqrt{(x_m-x_0)^2+(y_m-y_0)^2} \times0.2 \,,
\end{equation}
where $x_m$ and $y_m$ are the x and y coordinates of the peak of the density plot and $x_0$ and $y_0$ are the coordinates of the centre of the plot (SZ peak). We multiply the distance formula by the size of 1 pixel in Mpc (0.2), to obtain CS in the projected Mpc.

\subsubsection{Asymmetry Parameter}\label{sec3.1.2}
Another way to identify the morphology and dynamical state of a galaxy cluster is to study and classify the asymmetry of the cluster. The morphology of astronomical systems is closely related to their symmetry. If a cluster is asymmetric, then it has been disturbed in some way and formed substructures~\citep{2013MNRAS.436..275W}.
An efficient way to estimate the asymmetry of a cluster is to flip the density map horizontally and vertically and then compare it to the original image \cite{Conselice_2000}. The asymmetry parameter is calculated using the following formula~\citep{Conselice_2000}:
\begin{equation}
A^2=\frac{\sum (I_0 -I_\phi)^2}{\sum(2I_0^2)}\,,
\end{equation}
where $I_0$ is the original image and $I_\phi$ is the image that has been flipped horizontally and~vertically.

\subsection{X-ray Morphology}\label{sec3.2}
In this section, we show the calculations for the three main X-ray morphological parameters to further study the cluster dynamics and compare our results with those available in the literature. Following the work of~\citet{2008ASPC..399..375S},~\citet{2006MNRAS.373..881P}, and~\citet{1995ApJ...452..522B}, we calculate the power ratio, centroid shift value, and concentration parameter of~J0019.

\subsubsection{Concentration Parameter}\label{sec3.2.1}
Relaxed clusters generally host luminous cool cores in their centres whereas merging clusters show core disturbance~\citep{1984Natur.310..733F}. The concentration parameter is defined as the ratio of the X-ray integrated fluxes in the cluster core and the larger-scale region of the galaxy cluster~\citep{2008ASPC..399..375S}. The concentration parameter $c_{SB}$, following~\citet{2008ASPC..399..375S}, is calculated in two circular regions; a core radius of 100 kpc and an outer radius of 500 kpc using:
\begin{equation}
c_{SB} = \frac{S <(100  \rm~ kpc)}{S<(500 \rm~  kpc)},
\end{equation}
where $S$ is the X-ray surface brightness centred on the X-ray peak, within a particular radius (i.e., 100 kpc and 500 kpc).~\citet{2010ApJ...721L..82C} classifies a cluster as relaxed if $c_{SB}$ > 0.2 and disturbed if $c_{SB}$ < 0.2.

\subsubsection{Centroid Shift}\label{sec3.2.2}
The centroid shift method is the most robust X-ray morphology estimator, as it is most sensitive to the dynamical state while being least sensitive to image quality~\citep{2006MNRAS.373..881P}. The centroid shift is defined as the distance between the X-ray peak and the centre of mass, in units of aperture radius, $R_{ap}$, and is calculated using the following equation:
\begin{equation}
w=\left[\frac{1}{N-1} \sum (\Delta_i - \langle \Delta \rangle)^2\right]^{1/2} \times \frac{1}{R_{ap}}
\end{equation}
where $\Delta_i $ is the distance between the X-ray peak and the centroid of the $i${th} aperture and $\Delta$ is the mean of the sample of $\Delta_i$s.

This equation calculates the offset ($\Delta_i $) between the X-ray peak and the centroid using a series of concentric circles. The centroid of the $i${th} aperture is the centre of mass within the $i${th} aperture. A sample of $\Delta_i $s are computed within
a circular aperture of radius $R_{ap}$, ranging from $R_{ap}$ = 500 kpc to 0.05 $\times R_{ap}$, decreasing in steps of 5$\%$, as seen in~\mbox{\citet{2006MNRAS.373..881P}}.~\citet{2013A&A...549A..19W} classifies a cluster as relaxed if $w$ < 0.01 and disturbed if $w$ > 0.01.


\subsubsection{Power Ratios}\label{sec3.2.3}
The power ratio probes the underlying mass distribution to analyze the substructure of a cluster. It is calculated using a multipole decomposition of the potential of the two-dimensional projected mass distribution within a certain aperture~\citep{1995ApJ...452..522B}. Instead of mass, it is applied to the X-ray surface brightness. The third-order power ratio is one of the best indicators of the dynamical state of a cluster~\citep{doi:10.1080/21672857.2013.11519713}. The general m-order power ratio (where \emph{m} > 0) is defined as:
\begin{equation}
P_m = \frac{1}{2m^2R_{ap}^{2m}}(a_m^2+b_m^2)  \;\;\; {\rm and}
\end{equation}
\begin{equation}
P_0=a_0 {\rm ln}(R_{ap})^{2}
\end{equation}
where $a_0$ is the total intensity within aperture $R_{ap}$.
$a_m$ and $b_m$ are generic moments in polar coordinates, $R$ and $\phi$, given by:
\begin{equation}
a_m(r) = \int_{R'\lesssim R_{ap}} S(x') (R')^mcos(m\phi') d^2x'
\end{equation}
\begin{equation}
b_m(r) = \int_{R'\lesssim R_{ap}} S(x') (R')^msin(m\phi') d^2x'
\end{equation}

The $P_2$ quadrupole power represents the cluster ellipticity, $P_3$ represents the bimodal distribution, and $P_4$ is similar to $P_2$, but has power on much smaller scales. $P_3$ is most suitable to identify the presence of substructures or asymmetries~\citep{doi:10.1080/21672857.2013.11519713}. We make use of the normalized hexapole moment, $P_3/P_0$, to provide a clear measure of substructure calculated in an aperture of radius, $R_{ap}$ = 500 kpc, centred on the X-ray cluster centroid as seen in~\citet{1995ApJ...452..522B}. Based on previous studies \cite{2013A&A...549A..19W}, a cluster is disturbed if $P_3/P_0 > 10^{-7}$.

\subsection{Results and Comparison with the Literature}\label{sec3.3}
To compare our optical results to the literature, the above methods were applied to a sample of 98 galaxy clusters found in~\citet{2013MNRAS.436..275W}, with known cluster dynamical states.
With the set of results obtained from this sample, we can categorize clusters as relaxed or unrelaxed based on the criteria shown in Figure~\ref{fig4}.

We find that galaxy clusters with an asymmetry value, A$^2 > 0.1$ or a centre shift value, CS > 0.4 are all dynamically disturbed. Galaxy clusters with centre shift values, CS < 0.4, and asymmetry values < 0.1 could be classified as either relaxed or disturbed.

Applying the methods from Sections \ref{sec3.1.1} and \ref{sec3.1.2} on the density map of J0019~\mbox{(\ref{fig4})},\linebreak we obtain a centre shift value, CS = 0.20 $\pm$ 0.16 Mpc, and an asymmetry value,~\linebreak~\mbox{$A^2$~=~0.068~$ \pm$ $4.9\times10^{-4}$}.

The statistical uncertainties on the above parameters are estimated by Monte Carlo simulations in which we add Gaussian distributed noise to the galaxy photometry. The density map is recreated at each Monte Carlo step and after 1000 iterations, the standard deviation of the sample is taken as an estimate of the uncertainty.

\begin{figure}[H]
\includegraphics[width=\columnwidth]{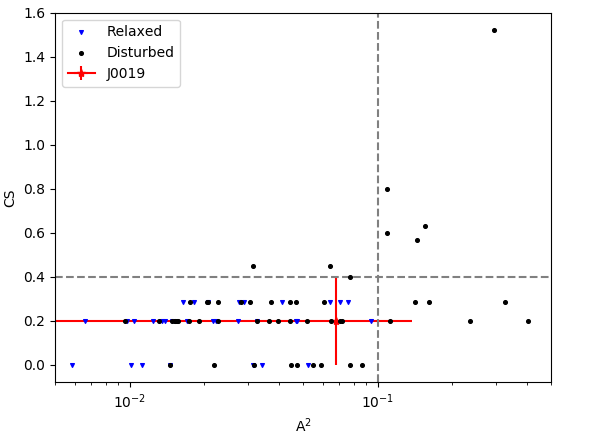}
\caption{ Plot of asymmetry parameter vs. centre shift for a sample of 98 clusters from 
\protect\citet{2013MNRAS.436..275W} with known cluster dynamical states. We find that all clusters with an asymmetry value greater than $\sim$0.1 (denoted by the vertical dashed line) are
unrelaxed. We also find all clusters with a centre shift value greater than $\sim$0.4 (denoted by
the horizontal dashed line) are unrelaxed. An asymmetry parameter less than $\sim$0.1 and
a centre shift value less than 0.4 show a mixture of relaxed and unrelaxed clusters. CS is in Mpc. The red cross represents J0019. \label{fig4}}
\end{figure}


The X-ray image has been smoothed with a Gaussian kernel of smoothing length 4~pixels for calculations. The following results were obtained: $P_3/P_0 = (8.93\pm 0.4)\times10^{-8}$, $w=0.03 \pm 0.01$, $c_{SB}= 0.13\pm 0.002$.
The analysis methods in Sections \ref{sec3.1.1} and \ref{sec3.1.2} 
have previously~\citep{2013A&A...549A..19W} been applied to a sample of 80 galaxy clusters observed with the XMM-Newton telescope and have classified a cluster as disturbed when $P_3/P_0 > 10^{-7}$ and $w > 0.01$. Another study~\citep{2010ApJ...721L..82C} also characterized cluster substructures based on the power ratios, centroid shift, and X-ray brightness concentration parameter estimated from Chandra X-ray images. They define a cluster to be dynamically disturbed if its morphological parameters satisfy the following conditions: $P_3 /P_0 > 1.2\times 10^{-7}, c_{SB} < 0.2$ and $w > 0.012$.
Two of the estimated morphological parameters ($c_{SB}$ and $w$) satisfy the above conditions, classifying J0019 as a dynamically disturbed cluster with $P_3/P_0$ falling~$\sim3\sigma$ away from the threshold value of $10^{-7}$.

The statistical uncertainties on the morphological parameters are estimated from a set of 1000 Monte Carlo simulations in which Poisson/shot noise is added to the observational XMM X-ray data. Uncertainties are estimated from the standard deviation of a sample of 1000 simulated images with added noise.

To verify our calculations of the three morphological parameters, we use XMM-Newton data on a known merging cluster, ACT-CL J0528.8$-$3927, and compare the derived morphological parameters to the known literature values found in~\citet{2013A&A...549A..19W} and~\citet{2017ApJ...846...51L}. All estimated values obtained by our code lie within $1\sigma$ of the literature values.

\subsection{Optical Redshift Distribution}\label{sec3.4}
The redshift distribution of 17 spectroscopically confirmed SDSS DR16~\citep{2020ApJS..249....3A} cluster member galaxies are used to gauge any disturbance in the cluster. This distribution is shown in Figure \ref{velocity} where there is an indication of bimodal structure in the histogram.

\begin{figure}[H]
\includegraphics[width=10cm]{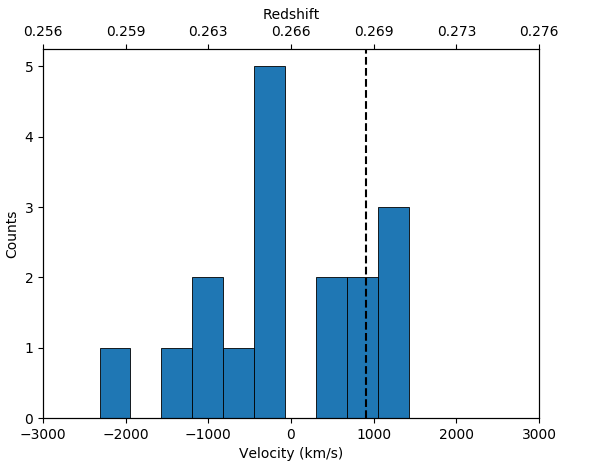}
\caption{Histogram showing the redshift distribution for 17
spectroscopically confirmed cluster members within $R_{500}$ from SDSS, DR16. Here, $v = 0$ is defined as the cluster systemic redshift of $z = 0.266$. The bin
width is 350 km/s. The vertical black dashed line shows the velocity of the BCG of the main component.}
\label{velocity}
\end{figure}

The standard deviation of the redshift distribution corresponds to the velocity dispersion, $\sigma_v$. Following~\citet{10.1093/mnras/stt049}, we use the calculated velocity dispersion and the galaxies-based scaling relation to determine $M_{200}$ and $R_{200}$, using a value of $h = 0.7$ in the~\citet{10.1093/mnras/stt049} equation. For comparison with other results, we convert our M$_{200}$ estimates into M$_{500}$, following the appropriate relation given in~\citet{10.1111/j.1745-3933.2008.00537.x}. The quantities M$_{500}$ and R$_{500}$ have been rescaled from M$_{200}$ and R$_{200}$ assuming a concentration parameter, $c = 3$. We integrate a NFW profile~\citet{1997ApJ...490..493N} and interpolate to determine M$_{500}$ and R$_{500}$. We estimate uncertainties on all cluster properties by bootstrap resampling 1000~times.

From the mean redshifts of the components, we find
a mean velocity at $z = 0.265$. The velocity dispersion\endnote{\label{note1}Corrected for the cosmology in this paper.}, $\sigma_v$, is 810 $ \pm~160$ km/s. We obtain a dynamical mass of~\linebreak~\mbox{M$_{200}$ = (4.1$~ \pm ~1.8) \times10^{14}M_\odot$} and M$_{500}$ = 2.5$~ \pm ~1.2 \times10^{14} M_\odot$.

\section{Discussion}\label{sec4}
It is believed that diffuse radio sources are associated with dynamically disturbed clusters. We can conclude that J0019 is undergoing a merger event through the detection of the radio halo shown in~\citep{2021MNRAS.504.1749K}. A radio halo in J0019 would require a significantly substantial merger to power it. Previous studies~\citep{2013A&A...549A..19W} have classified a cluster as disturbed when $P_3/P_0 > 10^{-7}$ and $w > 0.01$. The authors of
~\citep{2010ApJ...721L..82C} also characterized cluster substructures based on the power ratios, centroid shift, and X-ray brightness concentration parameter estimated from Chandra X-ray images. They define a cluster to be dynamically disturbed if its morphological parameters satisfy the following conditions: $P_3 /P_0 > 1.2\times 10^{-7}$, \mbox{$c_{SB} < 0.2$}, and $w > 0.012$.
The morphological estimators for J0019 (X-ray concentration and X-ray centre shift values), classify J0019 as a merging cluster and the X-ray $P_3/P0$, optical centre shift, and optical asymmetry values are inconclusive. X-ray morphology parameters (particularly the power ratios) are largely insensitive to substructures along the line of sight. Spectroscopy of more cluster members are needed to gauge any disturbed morphology in this direction.

The optical density map shows a displacement (0.2 Mpc) between the ACT cluster SZ peak and the peak in the density map. Moreover, the density map is fairly elongated and axisymmetric. The density map shows two components, one along a northwestern direction (top) and another along a southeastern direction (bottom). This is inferred through the `tails' or `trails' seen in these regions of the density map.

The radio halo in the left panel of Figure~\ref{sfig1} roughly follows the thermal gas in the X-ray image (right panel of Figure~\ref{sfig1}) as seen in the level edge present at the bottom of the diffuse structure and its overall shape in both images. The thermal gas and diffuse emission follow an elongated shape with an extended bottom left feature which hosts the BCG. The X-ray image orientation of the extended bottom left feature indicates that gas in the subcluster is undergoing ram pressure stripping as it interacts with the main cluster component.

Furthermore, there is a velocity offset\textsuperscript{\ref{note1}} of $\sim$950 km/s
between the mean velocity of the cluster and that of the BCG, further indicating that J0019 is a disturbed cluster.
There is also a discrepancy between the SZ mass ($M_{SZ,500}$), the X-ray estimated mass ($M_{X,500}$), and dynamical mass estimated from the optical redshift distribution ($M_{\rm{dyn},500}$). $M_{X,500}$ and $M_{SZ,500}$ lie $>3\sigma$ away from $M_{\rm{dyn},500}$, suggesting that the two components, northwest and southeast of the cluster centre, have already merged and J0019 is now observed in a post-merger phase. The gas traced by X-ray and SZ emission is completely disturbed during a previous collision, showing an enhanced X-ray emission and hence, a \mbox{T$_{X}-M_{X}$ overestimation}.

The angular resolution and short baselines of the MeerKAT telescope allow us to investigate both low surface brightness diffuse emission as well as compact radio sources.
The radio contours overlaid on the optical image are shown in Figure~\ref{fig5} where 10 compact radio sources are identified within a radius, R = 500 kpc. {The compact source properties are listed in Table} \ref{tab1}.

Only one of the radio sources is found in the Faint Images of the Radio Sky at Twenty-Centimetres survey (FIRST~\citep{1995ApJ...450..559B}). This source is labelled S5 and has an integrated flux density, S = 0.56 mJy. Source S3 in Figure~\ref{fig5} is associated with the BCG of the main cluster component and S4 is associated with the BCG of the infalling subcluster. This is inferred from the optically visible BCGs. There are two BCGs that are spatially separated, as seen in the DES image in Figure \ref{fig5}. These BCGs are also separated in velocity space as shown in Figure \ref{velocity}. This provides support for the existence of two distinct galaxy populations.


Moreover, an arched structure is found within the radio halo in projection. The optical image (Figure~\ref{fig5}) shows no optical counterpart at the centre of this arched structure, S1, suggesting that it is embedded in the diffuse structure. There are, however, two optical galaxies at either end of the bent radio structure. Spectroscopy of all cluster members are needed to gauge if these galaxies are related to the radio emission or not.

\begin{figure}[H]
\includegraphics[width=12cm]{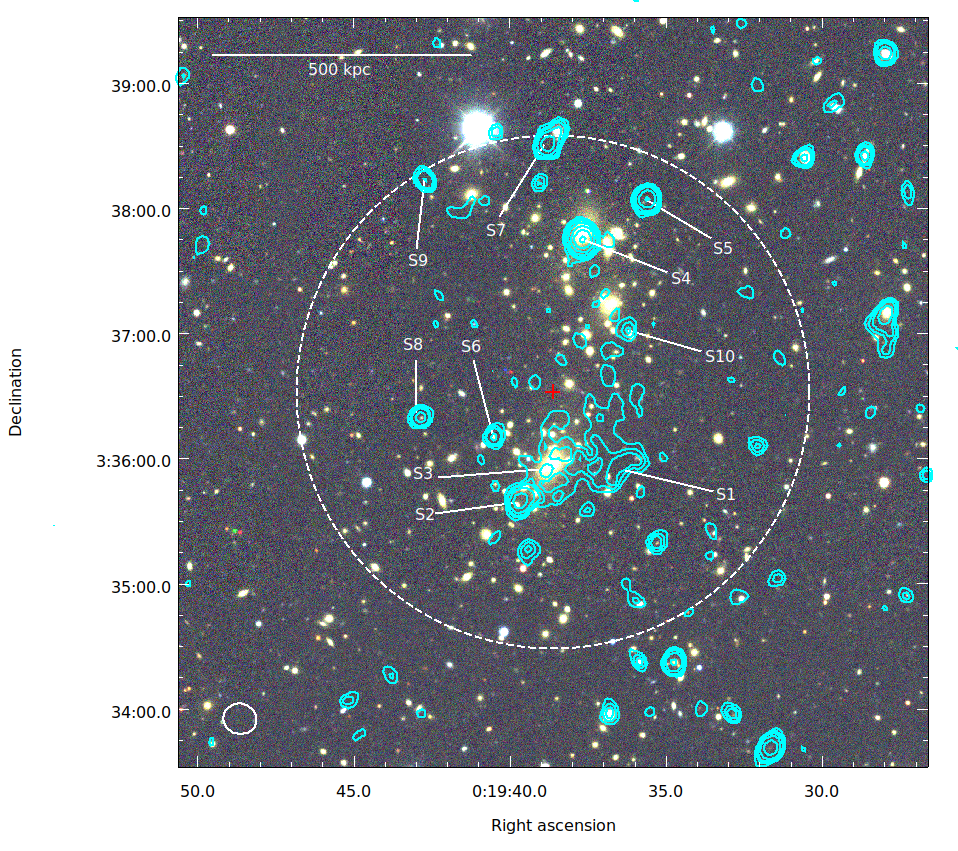}
\caption{Optical $gri$-band DES image with MeerKAT contours overlaid. Contours are without short baselines and start at [3, 4.5, 6, 10, 20, 40, 80]~$\times$ $\sigma$ rms where the rms is the full resolution central noise ($\sim$18.1 $\upmu$Jy/beam). The white dashed circle represents $R_{500 kpc}$. The radio image beam is shown as the white ellipse in the lower left corner (7.9$^{\prime\prime}$ $\times$ 7.2$^{\prime\prime}$, p.a. 12$^{\circ}$) and the red cross marks the position of the ACT SZ peak. Individual radio galaxies are labelled from S1 to S10. \label{fig5}}
\end{figure}

\vspace{-6pt}

\end{paracol}
\nointerlineskip
\begin{specialtable}[H]
\widetable
\caption{Properties of 
cluster region radio sources. Source labels are shown in Figure \ref{fig5}. The R.A. and Dec. values are for the peak source emission in the radio map. Flux densities for the sources are measured in the $1.16~\rm{GHz}$ map. Spectroscopic redshifts, z$_{\rm{Spec}}$, are from SDSS DR16.
$^a$ C: compact; T: resolved with tailed emission.\label{tab1}}
\widefigure
\setlength{\tabcolsep}{2.9mm}\begin{tabular}{ccccccc}
\toprule
\multirow{2}{*}{\textbf{Source}}& \textbf{R.A.} &\textbf{Dec.}&\multirow{2}{*}{\textbf{Type} \textsuperscript{\emph{\textbf{a}}}}&\textbf{S}\textsubscript{{\textbf{1.1}\rm{\textbf{6GHz}}}}& \multirow{2}{*}{z\textsubscript{{\rm{\textbf{Spec}}}}}&\multirow{2}{*}{\textbf{Notes}}\\
&\textbf{(deg)}&\textbf{(deg)}&&\textbf{(mJy)}&&\\

\midrule
S1&4.90205&3.59912&T&0.30 $\pm$ 0.03& & No optical counterpart found\\
S2&4.91539&3.59455&C&0.67 $\pm$ 0.03&& \\
S3&4.91164&3.59829&C&0.04 $\pm$ 0.01&0.26904&Associated with BCG of the main component\\
S4&4.90705&3.62907&C&1.70 $\pm$ 0.03& 0.26415&Associated with BCG of infalling subcluster\\
S5&4.89830&3.63448&C&0.69 $\pm$ 0.02&& Detected in FIRST(0.56 mJy)\\
S6&4.91872&3.60286&C& 0.10 $\pm$ 0.01 &&\\
S7&4.91153&3.64177&T& 0.66 $\pm$ 0.03&&\\
S8&4.92836&3.60541&C&  0.15 $\pm$ 0.02 &&\\
S9&4.92806&3.63719&C& 0.11 $\pm$ 0.01 &&\\
S10&4.90081 &3.61702 &C& 0.05 $\pm$0.01 &&\\
\bottomrule
\end{tabular}

\end{specialtable}
\begin{paracol}{2}
\switchcolumn

\vspace{-10pt}
\section{Conclusions}\label{sec5}
A low surface brightness radio halo was detected in J0019~\citep{2021MNRAS.504.1749K} with the MeerKAT telescope at L-band. The halo flux density estimated by~\citet{2021MNRAS.504.1749K} is~\linebreak~\mbox{$S_{1.16}$~=~$9.16 \pm 0.57$~mJy}. The calculated k-corrected 1.4 GHz radio power, $P_ {1.4GHz}$, is~\linebreak~\mbox{(1.0 $\pm$ 0.3) × $10^{24}$~W/Hz} and has been estimated by extrapolating the 610 MHz flux density to 1.4~GHz using a theoretically motivated spectral index of $\alpha$ = $-$1.3 $\pm$ 0.4. In this study, we performed a dynamical state analysis of J0019 using both XMM-Newton X-ray observations and optical DES photometry. The estimated morphology parameters are as follows: $P_3/P_0 = (8.93\pm 0.4)\times10^{-8}$, $w=0.03 \pm 0.01$, $c_{SB}= 0.13\pm 0.002$, $A^2$ = 0.068$~ \pm ~4.9\times10^{-4}$, and CS = 0.20 $\pm$ 0.16 Mpc.

Although the cluster is known to be involved in a merger event through the detection of the radio halo, the X-ray $P_3/P_0$ value, optical centre shift, and optical asymmetry values in the morphology analysis are inconclusive. The X-ray centroid shift and concentration values indicate a dynamically disturbed system.

The optical density map shows two cluster components: one along the northwestern and another in the southeastern direction. The BCG of the main component lies in the southern region and the BCG of the infalling cluster lies in the northern region as seen in the DES gri-band image and optical galaxy density map. Moreover, the density map shows that the cluster is elongated and axisymmetric.

{The bimodality detected in the velocity distribution and the offset of the BCG with respect to the mean velocity of the cluster suggests that the cluster merger has a component in the line-of-sight direction. However, the DES density maps, the MeerKAT radio halo, and XMM X-ray surface brightness shape shows an elongation in the north--south direction, which suggests that the collision also presents a velocity component contained in the plane of the sky and follows this north--south axis.}

With an elongated shape in the X-ray map, the clear bimodality and axisymmetric structures in the optical maps, two X-ray morphology parameters, the BCG offset by $\sim$950~km/s from the mean velocity, and the discrepancy between the SZ mass~\linebreak~($M_{SZ,500}$~= (10.2 $\pm~2.28)\times10^{14}M_\odot$), X-ray mass ($M_{X,500} \sim (6.5 \pm 1.3)\times10^{14} M_\odot$), and dynamical mass ($M_{\rm{dyn},500}$ = (2.5~$\pm~1.2)\times10^{14}M_\odot$), we clearly show that J0019 is a merging cluster and probably in a post-merging phase.


We can aim to obtain spectroscopy for the full field to identify more cluster members, infer cluster dynamics, and model how the merger event takes place in future work. An additional aim could be to study both the spectral index map of the diffuse structure and the polarized radio map. The energetics of the system can also be studied in future work with simulations.

\vspace{6pt}
\authorcontributions{Conceptualization, 
implementing methodology, interpretation of results, analysis, draft writing and editing, D.S.P.; resources and draft writing, D.J.T.; methodology, K.K., K.C.K., M.H.  and D.S.P.;  writing---review and editing, D.S.P., K.K., M.H., K.M., T.M., N.O., C.P., S.P.S.  and E.J.W.; supervision, K.M., K.K.  and M.H. All authors have read and agreed to the published version of the manuscript.}

\funding{The MeerKAT telescope is operated by the South African Radio Astronomy Observatory (SARAO), which is a facility of the National Research Foundation, an agency of the Department of Science and Innovation. DSP acknowledges funding from the South African Radio Astronomy Observatory and the National Research Foundation (NRF Grant Number: 96741). K.K. is supported by the New Scientific Frontiers grant of the South African Radio Astronomy Observatory. DT acknowledges support from the UK Science and Technology Facilities Council via grant ST/P006760/1. CP acknowledges funding from the European Research Council under ERC-CoG \mbox{grant CRAGSMAN-646955}.}

\institutionalreview{Not applicable.}

\informedconsent{Not applicable.}

\dataavailability{The radio data underlying this article will be shared on reasonable request to the corresponding author. XMM Newton observations and optical DES images and photometry are publicly available.}

\acknowledgments{%
The authors thank the anonymous referees whose comments have greatly improved the manuscript, and B. Partridge for useful comments on the original version.
This research made use of the Astropy, 13 (Astropy Collaboration et al. 2013, 2018), NumPy (Harris et al. 2020), and SciPy
(Virtanen et al. 2020) Python packages. Astropy is a community-developed core Python package for Astronomy. The Common Astronomy Software Applications (CASA) package is developed by an international consortium of scientists based at the National Radio Astronomical Observatory (NRAO), the European Southern Observatory (ESO), the National Astronomical Observatory of Japan (NAOJ), the Academia Sinica Institute of Astronomy and Astrophysics (ASIAA), the CSIRO division for Astronomy and Space Science (CASS), and the Netherlands Institute for Radio Astronomy (ASTRON) under the guidance of NRAO.}

\conflictsofinterest{The authors declare no conflict of interest.}








\end{paracol}
\printendnotes[custom]
\reftitle{References}





\end{document}